\documentclass[final,5p,twocolumn,times,number,compress]{elsarticle}

\usepackage{dcolumn,bm}
\usepackage{amsmath,amssymb,amsfonts}

\newcommand{\eps}{\epsilon}
\newcommand{\z}{\zeta}
\newcommand{\tends}{\rightarrow}
\newcommand{\Li}{\mathrm{Li}}
\newcommand{\kp}{\kappa}

\journal{Physics Letters A}

\begin{document}

\begin{frontmatter}

\title{Finite-size effects on the Bose-Einstein condensation critical temperature in a harmonic trap}

\author{J.M.B.~Noronha}
\ead{jnoronha@por.ulusiada.pt}
\address{Universidade Lus\'{\i}ada - Norte (Porto), Rua Dr. Lopo de Carvalho, 4369-006 Porto, Portugal}

\begin{abstract}
We obtain second and higher order corrections to the shift of the Bose-Einstein critical temperature due to finite-size effects. The confinement is that of a harmonic trap with general anisotropy. Numerical work shows the high accuracy of our expressions. 
We draw attention to a subtlety involved in the consideration of experimental values of the critical temperature in connection with analytical expressions for the finite-size corrections.
\end{abstract}


\end{frontmatter}

\section{Introduction}

In the first realizations of Bose-Einstein condensation (BEC) in the laboratory \cite{Anderson1995,Bradley1995,Davis1995} and in many experiments ever since, the Bose gas is trapped in a potential that can be considered as parabolic to a very good approximation. In the thermodynamic limit, within the ideal gas approximation, the critical temperature for such a system is given by $k_BT_0=(N/\zeta(3))^{1/3}\hbar \tilde\omega$, where $\tilde\omega =(\omega_x\omega_y\omega_z)^{1/3}$ is the geometric mean of the trap frequencies and all the other symbols have their usual meaning (see e.g. \cite{Dalfovo1999}). Soon after the first experiments, corrections to this expression, $\Delta T_c\equiv T_c-T_0$, were found. On the one hand, experiments do not take place in the thermodynamic limit. Hence, finite-size corrections 
are required. On the other hand, the gases are not ideal, having a non-vanishing scattering length. Hence, interaction effects must be taken into account.

The first order shift $\Delta T_c$ due to interactions was determined analytically early on in \cite{Giorgini1996} within a mean-field approximation, in the form of a linear term in the scattering length. Higher order corrections followed in several works 
\cite{Houbiers1997,ArnoldTomasik2001,Metikas2004,Zobay2004,ZobayMetikasKleinert2005,DavisBlakie2006,Briscese2013,Bronin2013,Haldar2014,Castellanos2015,Sergeenkov2015}, both numerical and analytical. Expansions for $\Delta T_c$ in powers of the scattering length down to second order were determined, both within a mean-field approach \cite{Zobay2004,Smith2011b,Briscese2013,Castellanos2015,Sergeenkov2015} and accounting for critical correlations \cite{ArnoldTomasik2001}.

The first order finite-size induced shift was given in the isotropic case in \cite{GrossmannHolthaus1995a,GrossmannHolthaus1995b} and in the general anisotropic case in \cite{KetterlevanDruten1996} as
\begin{equation}
\Delta T_c=-\frac{\z(2)}{2\z(3)}\frac{\hbar\bar{\omega}}{k_B}\; ,
\label{classic}
\end{equation}
where $\bar{\omega}=(\omega_x+\omega_y+\omega_z)/3$.
More recently, a higher order result was given in \cite{Jaouadi2011} (see also \cite{Noronha2015,Jaouadi2015}). 
This result relies on the local density approximation, in which the discrete energy levels of the finite system are approximated by a continuum, therefore requiring that the typical thermal energy at the transition be much greater than the typical inter-level spacing (for example, in the case of an isotropic harmonic trap, $k_BT_c\gg \hbar\omega$), i.e., the thermodynamic limit. Moreover, in order to overcome the vagueness (or non-point-like character) associated with the critical temperature of the finite system, we believe it would be useful to consider an explicit physical criterion for this critical temperature, related for example to the condensate fraction or the specific heat, when going to the level of detail of higher-order corrections \cite{Noronha2015}.

Strictly speaking, a finite-size correction to $T_0$ is an ill-defined concept when taken on its own
because the effect of finite size is to spread out the phase transition from a point to a narrow temperature interval. The first order correction (\ref{classic}) is typically  extracted from a high temperature finite-size expansion of the number of particles, which takes into due account the discreteness of the energy levels and which can be obtained in several ways \cite{KetterlevanDruten1996,KirstenToms1996dos,Haugerud1997,Haugset1997}. If one attempts to find a second order correction from this expansion, the absence of a true critical temperature makes itself noticed: the next order term in the expansion is divergent at the critical point, ultimately implying the non-existence of BEC as a sharp, mathematically defined phase transition in finite systems. It follows that the first order corrected $T_c$ must not be taken too seriously. It merely provides a reference value for signaling the transition.

In experimental work where the BEC critical temperature is measured \cite{Ensher1996,Gerbier2004,Smith2011a,Xiong2013}, the expression generally quoted for purposes of comparison with theory, namely for splitting off finite-size effects from interaction effects, is the one in (\ref{classic}). Now, as mentioned above, this expression should not be taken at face value. Thus, there is the possibility that a misinterpretation of the finite-size related shift can lead to a bias in the reported values of the interaction induced shift.
It would be of interest to make this matter clearer. 
What is actually measured 
in experiments
is the number of particles, ground state fraction, trap frequencies and temperature. It is by performing some polynomial fit to a plot involving these quantities that an experimental value for $T_c$ is usually extracted \cite{Ensher1996,Gerbier2004,Smith2011a}. In the landmark experiment reported in \cite{Gerbier2004} the fit is performed in the region where the condensate fraction ``noticeably starts to increase''. Condensate fractions as low as about $1\%$ could be measured in this experiment. If 
lower condensate fractions could be measured,
higher critical temperatures would have been obtained, even rising above $T_0$ for sufficiently small condensate fractions. This is because for finite systems the condensate fraction is not zero for temperatures above the critical region. It is just very small. This fact becomes more conspicuous for low particle numbers. 
Another major experiment in what concerns high precision measurements of $T_c$ is reported in \cite{Smith2011a}. Here, very much the same comments apply. In this case, condensate fractions as low as $0.1\%$ could be detected. The authors overcome the problem of isolating interaction from finite-size corrections by performing differential measurements with reference to a standard value of the scattering length. Nevertheless, as recently pointed out \cite{Briscese2013}, this assumes that finite size and interaction effects are independent. At second order, it might not be the case.

Our aim in the present work is to obtain  
higher-order finite-size corrections to the critical temperature of a Bose gas in a general harmonic trap. 
To do this in a meaningful way, 
which at the same time can connect to experimental procedures, 
we overcome the non-existence of a true critical temperature by asking instead for the temperature $T_{\kappa}$ at which the condensate fraction has a given small value $N_{\textrm{gr}}/N=\kappa$, $\kappa\ll 1$. Other criteria could be used, like defining $T_c$ by the maximum of the specific heat or the inflection point of the $N_{\textrm{gr}}(T)$ curve; but the one we adopt here is probably the most useful because it uses the condensate fraction and it is very simple.
From the well known bulk behaviour of the condensate fraction in the BEC regime, $N_{\textrm{gr}}/N=1-(T/T_0)^3$, we have in the thermodynamic limit $T_{\kappa}/T_0=(1-\kappa)^{1/3}$. For $\kappa\tends 0$, this yields $T_{\kappa}\tends T_0$. We will provide finite-size corrections to $T_{\kappa}$ down to third order. Stopping at second order is not accurate enough in some circumstances, as detailed below. Our approach preserves all the finite-size characteristics of the system, with no approximations involved. The information on the discrete structure of the energy levels is carried in the expansions (\ref{condfraction}) and (\ref{t_k}) below. Finally, we note that our expressions are also valid (and highly accurate) for $\kappa$ not small, i.e., deep into the BEC regime.

\section{Finite-size corrections}

Let 
$x=\beta\hbar\bar{\omega}=\hbar\bar{\omega}/(k_BT)$ 
and $\epsilon=(E_{\textrm{gr}}-\mu)/(\hbar\bar{\omega})$. $x$ is a rescaled inverse temperature and $\epsilon$ can 
be looked at as a rescaled chemical potential. We define the anisotropy vector $\bm{\lambda}=
(\lambda_1,\lambda_2,\lambda_3)=(\omega_1,\omega_2,\omega_3)/\bar{\omega}$. Using grand-canonical statistics, the number of particles $N$ of an ideal Bose gas in this trap is given by
\begin{equation}
N
=\sum_{\mathbf{n}}\left[ e^{\beta(E_{\mathbf{n}}-\mu)}-1\right] ^{-1} 
=\sum_{\mathbf{n}}\sum_{k=1}^{\infty}e^{-kx(\bm{\lambda}\cdot\mathbf{n}+\epsilon)} \; .
\label{Nnew}
\end{equation}
The sum in $\mathbf{n}$ is over all single particle states, of energy $E_{\mathbf{n}}=\sum_{i=1}^3\left(n_i+1/2\right)\hbar\omega_i$, $n_i=0,1,2,\ldots$. Let $\lambda=(\lambda_1\lambda_2\lambda_3)^{1/3}$. The usual bulk result for $N$, which is exact in the thermodynamic limit, reads in our variable
$x^3N=\Li_3(e^{-x\epsilon})\lambda^{-3}$ if  $T\geq T_0$ and $x^3N=
x^3N_{\textrm{gr}}+\z(3)\lambda^{-3}$ if $T<T_0$
(where $x^3N$ is the quantity that remains finite in the thermodynamic limit, as opposed to $N$). $\Li_3$ is the polylogarithm of index 3, with the property $\Li_3(1)=\z(3)$. Define $x_0=\hbar\bar{\omega}/(k_BT_0)=(\z(3)/N)^{1/3}\lambda^{-1}$. As we approach the thermodynamic limit in the usual way ($N\omega_i^3$ kept fixed) we have $x_0\tends 0$, or for any fixed temperature, $x\tends 0$. $x$ and eventually $x_0$ will be our expansion parameters. In the BEC regime, we have in addition (still in the thermodynamic limit) $N_{\textrm{gr}}=1/(\epsilon x)$, from where we see that $\epsilon$ scales as $x^2$.

What we need is an expansion for $N$ that contains the finite-size corrections and that is valid \textit{throughout} the critical region. This can be achieved by applying a Mellin-Barnes transform to the exponential inside the  $k$ summation in (\ref{Nnew}), as indeed was done before in \cite{KirstenToms1996MB}. The same procedure was also applied to a Bose gas subject to other confinements \cite{Toms2006,NoronhaToms2013}. 
An expansion is obtained by solving a contour integral in the complex plane using the theorem of residues. In this case, the Riemann and three-dimensional Barnes zeta functions, here denoted $\z(\alpha)$ and $\z_B(\alpha,\epsilon|\bm{\lambda})$ respectively, make their appearance. Knowledge of the residues at the poles of these functions is required. We refer the reader to \cite{KirstenToms1996MB} for details of the procedure.
$\z_B$ is a multi-dimensional generalization of the Hurwitz zeta function, which was studied in depth by Barnes in \cite{Barnes1904} (see also \cite{Kirsten:book}). In \cite{KirstenToms1996MB} the expansion for $N$ was calculated to subleading order. However, for our purposes we need also the third and fourth terms. The calculation of the third term, in particular, is more involved due to the existence of a double pole, requiring the knowledge of the finite part at the $\alpha =1$ pole of $\z_B(\alpha,\epsilon|\bm{\lambda})$, not only its residue. 
Specifically, below we need the quantity $b_0(\bm{\lambda})$ defined in the following way. Let $a_0(\epsilon|\bm{\lambda})$ be the finite part at the $\alpha=1$ pole of $\z_B(\alpha,\epsilon|\bm{\lambda})$. Then $b_0=\lim_{\epsilon\tends 0}(a_0(\epsilon|\bm{\lambda})-\epsilon^{-1})$, i.e., $a_0(\epsilon|\bm{\lambda})=\epsilon^{-1}+b_0+\mathcal{O}(\epsilon)$. $b_0$ is a function of $\bm{\lambda}$ only. 
We obtain the expansion
\begin{multline}
N=\frac{\z(3)}{\lambda^3}x^{-3}+\frac{3-2\epsilon}{2\lambda^3}\z(2)x^{-2}
+\biggl[ a_0(\eps|\bm{\lambda})
\\
\left. -\frac{9+(\lambda_i\lambda_j)-18\eps+6\eps^2}{12\lambda^3}\ln x\right] x^{-1}
-\frac{1}{2}\z_B(0,\eps|\bm{\lambda}) +\mathcal{O}(x) \; ,
\label{Nexpansion}
\end{multline} 
where  
we have adopted the following notational conventions: $(\lambda_i\lambda_j)=\sum_{i,j=1\, (i<j)}^3\lambda_i\lambda_j=\lambda_1\lambda_2+\lambda_1\lambda_3+\lambda_2\lambda_3$ and $(\lambda_i^2\lambda_j)=\sum_{i,j=1\, (i\neq j)}^3\lambda_i^2\lambda_j$. 
The first two terms in (\ref{Nexpansion}) were given in \cite{KirstenToms1996MB}. From \cite{Barnes1904} we have that $\z_B(0,\epsilon|\bm{\lambda})=1/8+(\lambda_i^2\lambda_j)/(24\lambda^3)+\mathcal{O}(\eps)$.
The full asymptotic expansion for $N$ could easily be given, but it is not needed. 
 
Define the rescaled temperature $t=T/T_0=x_0/x$. In (\ref{Nexpansion}), change from the variables $N$, $x$ and $\eps$ to $x_0$, $t$ and $\epsilon$ by performing the substitutions $N=\z(3)/(\lambda x_0)^3$ and $x=x_0/t$. Equation~(\ref{Nexpansion}) gives us $\epsilon$ implicitly as a function of $x_0$ and $t$. Since $\epsilon=\mathcal{O}(x_0^2)$ for $t<1$, we solve for $\epsilon$ perturbatively by letting $\epsilon =a(t)x_0^2+b(t)x_0^3+c'(t)x_0^4\ln x_0+c(t)x_0^4+\cdots$ and find the coefficients $a(t)$, $b(t)$ \ldots 
Next we use the expression for the condensate fraction $N_{\textrm{gr}}/N=(e^{\epsilon x}-1)^{-1}/N$. In this expression, we change again to the variables $x_0$, $t$ and $\epsilon$ and substitute the newly found expansion for $\epsilon$. Expanding the resulting expression in powers of $x_0$ yields
\begin{multline}
\hspace{-0.35cm}
\frac{N_{\textrm{gr}}}{N}=\left( 1-t^3\right) -\frac{3\z(2)}{2\z(3)}t^2x_0-\frac{\lambda^3t}{\z(3)}\left[
b_0+\frac{9+(\lambda_i\lambda_j)}{12\lambda^3}(\ln t-\ln x_0)\right] \\
\times x_0^2
+\frac{\lambda^3}{\z(3)}\left[
\frac{\z(2)}{\z(3)}\frac{t^3}{1-t^3}+\frac{(\lambda_i^2\lambda_j)}{48\lambda^3}-\frac{7}{16}\right]x_0^3+\cdots\; .
\label{condfraction}
\end{multline}
This equation gives us the condensate fraction as a function of $t$ and $N$ (or $t$ and $x_0$). It is valid throughout the BEC regime and critical region. Note that the first two terms, $1-t^3$, are just the bulk result for $N_{\textrm{gr}}/N$ in the condensate region. 
We then set the condensate fraction at  $N_{\textrm{gr}}/N=\kp$ and solve (\ref{condfraction}) 
perturbatively, this time to find $t_{\kp}\equiv T_{\kp}/T_0$ as a function of $\kappa$ and $x_0$. This finally yields
\begin{multline}
t_{\kp}=A(\kp)+B(\kp)x_0+C(\kp)x_0^2+C'(\kp)x_0^2\ln x_0+D(\kp)x_0^3\\
+D'(\kp)x_0^3\ln x_0+\mathcal{O}(x_0^4\ln x_0) \; ,
\label{t_k}
\end{multline}
with the coefficients being given by
\begin{align*}
A(\kp)&=(1-\kp)^{1/3}\\
B(\kp)&=-\frac{\z(2)}{2\z(3)}\simeq -0.6842\\
C(\kp)&=\frac{1}{(1-\kp)^{1/3}}
\left[ \frac{\z(2)^2}{4\z(3)^2}-\frac{\lambda^3}{3\z(3)}\left( b_0+
\frac{9+(\lambda_i\lambda_j)}{36\lambda^3}\ln (1-\kp)\right)\right]\\
C'(\kp)&=\frac{9+(\lambda_i\lambda_j)}{(1-\kp)^{1/3}36\z(3)}\\
D(\kp)&=(1-\kp)^{-2/3}\left[ -\frac{\z(2)^3}{12\z(3)^3}+\frac{(\lambda_i^2\lambda_j)-21\lambda^3}{144\z(3)} 
+\frac{\z(2)}{6\z(3)^2}\right.\\
&
\left. \times
\frac{9+(\lambda_i\lambda_j)}{12}
\left( 1+\frac{1}{3}\ln (1-\kp) \right) +\frac{\z(2)\lambda^3}{6\z(3)^2}\left( b_0-2+\frac{2}{\kp}\right) \right] \\
D'(\kp)&=-\frac{\z(2)(9+(\lambda_i\lambda_j))}{36\z(3)^2(1-\kp)^{2/3}}\; .
\end{align*}
Note that since $x_0=(\z(3)/N)^{1/3}\lambda^{-1}$ this is an expansion in powers of $N^{-1/3}$. The leading term is just the bulk result for $t_{\kp}$. The subleading term is the well known first order finite-size correction to the critical temperature given in (\ref{classic}): $B(\kp)x_0=-\z(2)\hbar\bar{\omega}/(2\z(3)k_{B}T_0)$. The higher order terms are new. It is interesting to note that when $\kp\rightarrow 0$ the coefficient $D(\kp)$, unlike the other coefficients above, diverges (due to the $\kp$ in the denominator of the very last term). We return to this point in the next section. As mentioned above, this expansion is valid not only in the critical region ($\kp\ll 1$), but also throughout the BEC regime ($\kp$ not small).

In the isotropic case, $\lambda^3=1$, $(\lambda_i\lambda_j)=3$, $(\lambda_i^2\lambda_j)=6$ and, 
by writing $\zeta_B$ in terms of Hurwitz zeta-functions, it is easily seen that
 $b_0=\gamma-19/24\simeq -0.2145$, where $\gamma$ is Euler's constant.
 The $x_0^2$ and $x_0^3$ coefficients in (\ref{t_k}) are then given more simply as
\begin{align*}
C_i(\kp)&\simeq (0.5276-0.0924\ln(1-\kp))(1-\kp)^{-1/3}\\
C'_i(\kp)&\simeq 0.2773(1-\kp)^{-1/3}\\
D_i(\kp)&\simeq (-0.5306+0.3795\kp^{-1}+0.0632\ln(1-\kp))(1-\kp)^{-2/3}\\
D'_i(\kp)&\simeq -0.1897(1-\kp)^{-2/3} \; ,
\end{align*}
where we have used the numerical values of $\z(2)$ and $\z(3)$.
The subscript $i$ stands for``isotropic''. 

In order to use (\ref{t_k}) (or for that matter, any of the previous expansions to more than subleading order)  in the case of an anisotropic trap, we must be able to find $b_0$ in the general case.
From Barnes's work \cite{Barnes1904} (pp.398 and 404), it is easy to arrive at $b_0=\gamma_{31}(\bm{\lambda})$, where the $\gamma_{ij}$ are gamma modular forms, which Barnes gives quite generally in terms of contour integrals in the complex plane. Application to our case yields
\begin{multline}
b_0=\int_0^{\infty}dt\, \left[\prod_{r=1}^3\left( 1-e^{-\lambda_r t}\right)^{-1}-1-\frac{1}{\lambda^3t^3}-\frac{3}{2\lambda^3t^2}\right. \\ \left. 
-\frac{9+(\lambda_i\lambda_j)}{12\lambda^3t}e^{-t}\right] \; ,
\label{b0integral}
\end{multline}
where we have made use of the fact that $\lambda_1+\lambda_2+\lambda_3=3$, from the definition of $\bm{\lambda}$.
Table \ref{b0list} presents values of $b_0$ in a few illustrative cases. In the experiment by Gerbier \textit{et al} \cite{Gerbier2004}, the trap is cigar shaped with aspect ratio $47.5$. The widely used trap of Ensher \textit{et al} \cite{Ensher1996} is disc shaped with aspect ratio $\sqrt{8}$. We include these two shapes in the table. For a more complete table, see the supplementary material, where we give values of $b_0$ for axially symmetric traps with integer aspect ratios ranging from $1$ to $100$.
\begin{table}
\begin{tabular*}{\linewidth}{c@{\extracolsep{\fill}}D{.}{.}{4}D{.}{.}{4}}
\multicolumn{1}{c}{Aspect ratio} & \multicolumn{1}{c}{Disc shape} & \multicolumn{1}{c}{Cigar shape}\\
\hline
2 & 0.1991 & 0.2433\\
3 & 0.9605 & 1.1317\\
5 & 3.0403 & 3.5180\\
10 & 10.8558 & 11.4467\\
30 & 73.7461 & 55.4503\\
Gerbier \textit{et al} & & 102.0169\\
Ensher \textit{et al}&  0.8147 &
\end{tabular*}
\caption{\label{b0list} Values of $b_0$ for several trap shapes.}
\end{table}

\section{Discussion of results and conclusions}

\begin{figure}
\resizebox{\linewidth}{!}{\includegraphics{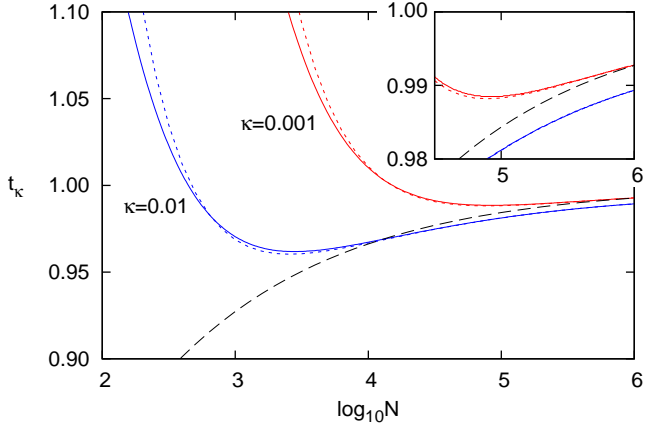}}
\caption{\label{fig:isotropic}(Colour online.) Rescaled temperature $t_{\kappa}=T_{\kappa}/T_0$ at which the condensate fraction is $\kappa$ as a function of $\log_{10}N$ for an isotropic trap. The solid lines are numerical and the accompanying dotted lines are analytical from Eq.~(\ref{t_k}). The $\kappa=0.01$ and $\kappa=0.001$ cases are set in blue and red colour, respectively.
The dashed line (in black) is the first order result from Eq.~(\ref{classic}). The inset zooms in on the large $N$ region.}
\end{figure}

In Fig.~\ref{fig:isotropic}
we plot $t_{\kappa}$ for $\kappa=0.01$ and $\kappa=0.001$ in the isotropic case. We plot both purely numerical results and our analytical results from (\ref{t_k}). Since (\ref{t_k}) is an expansion in powers of $N^{-1/3}$, its accuracy increases for larger $N$. For small $N$, the value of $\kp$ cannot be chosen too small. This is because in this situation we will have $t>1$ due to the spreading out of the phase transition, while we require $t\lesssim 1$. This can also be seen by looking at the coefficient of the $x_0^3$ term in (\ref{t_k}), $D(\kp)$. It contains a term with factor $1/\kp$ which becomes large if $\kp$ is very small. In fact, if we carry on with the expansion, it is seen that at every third term a new factor $1/\kp$ will appear. Thus, the terms in $x_0^3$, $x_0^4$ and $x_0^5$ contain the factor $1/\kp$, the terms in $x_0^6$, $x_0^7$ and $x_0^8$ contain the factor $1/\kp^2$ and so on. It follows that this expansion is valid only if $x_0^3/\kp\ll 1$, or equivalently, $\kp\gg 1/N$. This is a very reasonable condition. It corresponds exactly to temperatures very close to $T_0$ or below it, confirming in this way the original requirement $t\lesssim 1$ for the validity of the expansion. If $\kp\gg x_0^3$ just holds but the tighter condition $\kp\gg x_0^2$ (equivalently, $\kp\gg N^{-2/3}$) does not, then the expansion is valid but the term in $x_0^3$ is of the same order as the $x_0$ term (while the terms in $x_0^n$, $n>3$, will be of smaller order). Hence, it is very important to include it. This limiting situation happens for example for $N=10^3$ and $\kp=0.01$ or for $N=10^4$ and $\kp=0.001$. All this is very well corroborated by comparing with numerical results as can be seen in the figure. For medium to large $N$, the analytical and numerical curves are superimposed or hardly distinguishable, due to the high accuracy of (\ref{t_k}). For example, for $N\gtrsim 4\times 10^4$ and $\kappa=0.01$ or $N\gtrsim 2\times 10^5$ and $\kappa=0.001$ the error in $t_{\kp}$ from (\ref{t_k}) is less than $10^{-4}$ (in the isotropic case). As we lower $N$, the accuracy slowly decreases. As expected, this happens more quickly for $\kappa=0.001$. The broadening of the phase transition is observed in the rise of $t_{\kappa}$ for low $N$. Our approximation captures this behaviour.
For comparison, we plot the usual first order result, given by Eq.~(\ref{classic}).
We see it provides a useful reference value for the transition for $N\gtrsim 10^4-10^5$, even though it does not have a precise meaning. For lower values of $N$, the new corrections are particularly important. Roughly, in most situations, the new corrections should be important for $N\lesssim 10^5$. In such cases, if the interaction strength is small, it is quite possible that these effects become of the same order or even dominant over the interaction effects.

\begin{figure}
\resizebox{\linewidth}{!}{\includegraphics{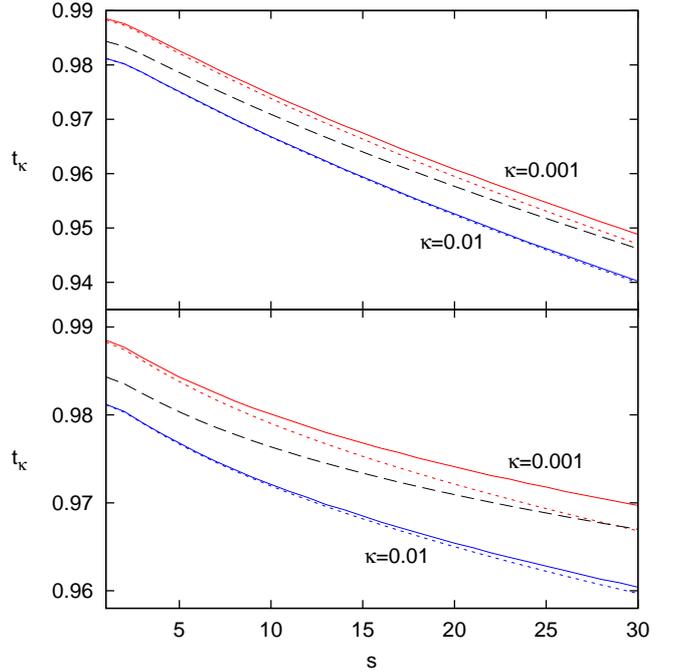}}
\caption{\label{fig:anisotropic}(Colour online.) Rescaled temperature $t_{\kp}$ as a function of the aspect ratio $s$ for axially symmetric disc shaped (upper plot) and cigar shaped (lower plot) traps. The several line patterns and colours have the same meaning as in Fig.~\ref{fig:isotropic}.}
\end{figure}

Anisotropy does not substantially modify the above analysis. In Fig.~\ref{fig:anisotropic}, 
we plot $t_{\kp}$ for $N=10^5$ and axially symmetric disc shaped and cigar shaped traps. $t_{\kp}$ is plotted as a function of the aspect ratio $s$. For both shapes, the anisotropy causes a decrease in $t_{\kp}$. This effect is more pronounced for the disc shape, which can be understood from the fact that the dependency of $\lambda$ (and hence of $x_0$) on the aspect ratio $s$ is stronger in the disc case. The effect seen in Fig.~\ref{fig:isotropic} where $t_{\kp}$ rises above $1$ for low $N$ is not seen in Fig.~\ref{fig:anisotropic} because anisotropy lowers the critical temperature without spreading out the phase transition, which remains sharp.
It can also be seen that the accuracy of (\ref{t_k}) slightly decreases with increasing anisotropy. This is expected and is due to the fact that $\lambda$ decreases with increasing $s$, which causes $x_0$ to increase. Naturally, this is more noticeable for $\kp=0.001$. It is also more noticeable in the cigar shape case. This should be due to the exact nature of the higher order terms ($x_0^4$ and higher) left out of our expansion (\ref{t_k}). All in all, it can be seen that (\ref{t_k}) is still quite accurate in most anisotropic situations (the exception being the highly anisotropic cigar shaped trap with very small $\kp$). As in the isotropic case, higher values of $N$ lead to higher accuracy. We also plot the first order result from (\ref{classic})  
in Fig.~\ref{fig:anisotropic}. Again, we see that for this number of particles, Eq.~(\ref{classic}) works well for providing a reference value for the transition.

Let us now take the experimental conditions of Smith \textit{et al} \cite{Smith2011a}: $N\simeq 10^5 - 10^6$ and a nearly isotropic trap. Condensate fractions as low as $\kappa \simeq 0.001$ could reliably be measured in this experiment. For $N=10^5$ and $\kappa=0.001$ we have $t_{\kappa}=0.9885$ from numerical calculations. Eq.~(\ref{t_k}) yields $t_{\kappa}=0.9883$ whereas the result from (\ref{classic}) 
yields $t_c\equiv T_c/T_0=0.9843$, which is lower than $t_{\kp}$ by 
$0.4\%$ of $T_0$. In Gerbier \textit{et al} \cite{Gerbier2004}, the trap is cigar shaped with aspect ratio $\sim 47.5$ and $N\simeq 1.5\times 10^5 -1.5\times 10^6$ at the transition. Condensate fractions as low as $\kappa\simeq 0.01$ could be measured. For this trap shape with $N=10^5$ and $\kappa=0.01$, we have $t_{\kappa}=0.953$ whereas (\ref{t_k}) yields $t_{\kappa}=0.952$. The first order result (\ref{classic}) 
yields $t_c=0.962$, which is higher than $t_{\kappa}$ 
by $1\%$ of $T_0$. For $N=10^6$ this difference reduces to $0.5\%$ of $T_0$. If the trap had a larger anisotropy, the difference would be larger. 
In order to obtain the shift $\Delta T_c$ due to interactions, the authors in \cite{Gerbier2004} subtracted the finite-size correction as given in (\ref{classic}) from their experimental values for $T_c$. 
If the authors had used the lowest reliably detected condensate fraction ($\kappa\simeq 0.01$) for defining an experimental critical temperature, the finite-size shift that should be subtracted would be $t_\kappa$ and not the one given in (\ref{classic}). However, the procedure for obtaining this temperature was actually more complex than that, involving linear fits to the plots of $N_{\text{gr}}$, $N$ and $T$ as functions of the trap depth. The main feature of this idea can perhaps be understood by thinking of a linear fit applied to the experimental points near the $\kappa=0.01$ region of the familiar $N_{\text{gr}}/N(T)$ curve. In this simplified version, the experimental $T_c$ would then be given by the point of intersection of the straight line of the linear fit with the horizontal ($T$) axes. Due to the complexity of the whole procedure, it is not clear exactly what finite-size correction should be subtracted. Nevertheless, we see that an error of the order of 1\% of $T_0$ (not more) could be involved in the determination of the interaction induced $\Delta T_c$ due to the use of eq.~(\ref{classic}) instead of a more precise expression.\footnote{Naturally, the experimental accuracy is also relevant here. In particular, a certain accuracy is necessary in order to observe higher-order finite-size effects. For example, for a number of particles $N\sim10^5$, an accuracy of about $0.5\%$ or less in the reported values of $t$ should suffice. This requirement seems to be quite realistic and, in particular, it seems to hold in the experiment of \cite{Smith2011a}.}

These examples illustrate the relevance of accurate analytical finite-size corrections, while suggesting the usefulness of better defined criteria for measuring the critical temperature, whenever finite-size effects play a role. In particular, in such cases
it would seem to us perhaps more relevant to talk about $t_{\kappa}$, the temperature at which the condensate fraction is $\kappa$, rather than \textit{the} critical temperature. For lower particle numbers, these considerations are even more important,
as can be seen in Fig.~\ref{fig:isotropic}.

On another note, the use of BEC experiments to probe Planck-scale physics has been suggested in the last few years (see \cite{Castellanos2012,Castellanos2014,Briscese2012} and references therein). The idea is that a quantum gravity effect could alter the single particle energy spectrum of the atoms in a harmonic trap, with a consequent shift in $T_c$. For this effect to show, we would ideally have a very weakly interacting gas and relatively small atom numbers\footnote{Some amount of interaction is necessary for thermal equilibrium, which could cast some doubt on the practicality of such experiment. Still, BECs of essentially ideal gases have been produced in several experiments, 
using Feshbach resonances. (See e.g. \cite{Roati2007}. In this particular experiment the shift in $T_c$ due to finite size is estimated to be about $2\%$, while the shift due to interactions is less than $0.001\%$, but thermal equilibrium cannot be assumed.)}. Finite-size effects would be crucial in such an experiment. In this context, the need has been recognized \cite{Briscese2012} for higher order finite-size corrections. We have given these corrections in the present work.   

This work took as a starting point a certain criterion for the BEC critical temperature of a finite system. As mentioned above, other physical criteria could be adopted. In principle, it should be possible to apply the techniques of the present work to these alternative criteria.

Finally, it would be interesting to study analytically the interplay between finite size and interaction effects. It would not be surprising if there is a second order correction cross term containing a dependency on both the finite size and the interaction strength.

\end{document}